\documentclass[prl,superscriptaddress,twocolumn,noshowpacs]{revtex4-1}
\usepackage{graphicx}
\usepackage{times}
\usepackage{enumitem}
\usepackage{amsmath,amssymb}
\newcommand{\be}{\begin{equation}}
\newcommand{\ee}{\end{equation}}
\newcommand{\bea}{\begin{eqnarray}}
\newcommand{\eea}{\end{eqnarray}}
\newcommand{\zt}{{\mathbb Z}_2}
\newcommand{\comment}[1]{}

\newcommand{\p}[1]{\langle #1\rangle}
\newcommand{\q}[1]{[[#1]]^{}}
\newcommand{\s}[1]{\{\{#1\}\}^{}}

\begin{document}

\title{Strong zero modes and eigenstate phase transitions in the XYZ/interacting Majorana chain}

\author{Paul Fendley\\
{\it\small All Souls College and Rudolf Peierls Centre for Theoretical Physics, 1 Keble Road, Oxford, OX1 3NP, United Kingdom}}
\vskip1in
\date{\today} 

\begin{abstract} 

I explicitly construct a strong zero mode in the XYZ chain or, equivalently, Majorana wires coupled via a four-fermion interaction. The strong zero mode is an operator that pairs states in different symmetry sectors, resulting in identical spectra up to exponentially small finite-size corrections.  Such pairing occurs in the Ising/Majorana fermion chain and possibly in parafermionic systems and  strongly disordered many-body localized phases. The proof here shows that the strong zero mode occurs in a clean interacting system, and that it possesses some remarkable structure  -- despite being a rather elaborate operator, it squares to the identity. Eigenstate phase transitions separate regions with different types of pairing.

\end{abstract} 
\maketitle
\paragraph{Introduction: eigenstate phase transitions --}

A quantum phase transition is by definition a property of the ground state, occurring where the ground-state energy is non-analytic in the couplings \cite{Subirbook}. A focus on the ground state and the low-lying excitations is natural, given that these dominate physical properties at low temperatures where quantum effects are most prominent. However, recent work on many-body localization (MBL) \cite{Husereview} has focused attention on the many interesting properties of the full energy spectrum.
One fascinating idea arising is an {\em eigenstate phase transition}, where properties of full spectrum undergo a sharp transition as couplings are varied \cite{HNOPS}. Such transitions need not occur concern the ground state; one proposal of Ref.\ \cite{HNOPS} is a change between types of statistics governing all energy-level spacings.

One substantial complication in understanding eigenstate phase transitions in the MBL context is that  properties of the full spectrum (e.g.\ the eigenstate thermalization hypothesis) are typically subtle. Complicating the analysis further is that strongly disordered couplings are typically needed to obtain the novel physics. Thus unfortunately the dominant tools utilized in these studies are intuition and numerical analysis. 

I discuss in this paper an eigenstate phase transition that has the virtues of being fairly easy to characterize, and tractable analytically. This transition is for a {\em strong zero mode}, an {\em operator} that commutes with the Hamiltonian, up to corrections that fall off exponentially in system size \cite{KitMajorana,PFpara,Jermyn}. I prove that a strong zero mode occurs in a system without disorder, with interesting consequences for the full energy spectrum.

The canonical example of a strong zero mode is in the Ising/Majorana quantum chain in the ordered phase with free boundary conditions on both ends. The strong zero mode pairs each state in the even-fermion number with one in the odd, requiring their energies are the same up to exponentially small finite-size corrections \cite{Pfeuty,KitMajorana}. 
It therefore results in a much stronger constraint on the energy spectrum that the constraint of (topological) order, where only pairing of ground states is necessary. This behavior is robust; the pairing persists when the couplings are allowed to vary spatially, as long as the strong zero mode operator remains normalizable (i.e.\ when acting on any normalizable state it gives another one). The strong zero mode and hence the pairing go away precisely at the quantum phase transition between spin/topological order and disorder, so in this case eigenstate and quantum phase transitions occur at the same coupling.

There have been suggestions that the strong zero mode does not survive the inclusion of interactions, unless the disorder is strong enough to create an MBL phase \cite{HNOPS,Bauer,Bahri}. This is true in some cases. The three-state Potts chain has in a trivially solvable limit a threefold analog of the pairing, but the strong zero mode disappears for arbitrarily small interactions, although the spin/topological order and a ``weak'' zero mode remain \cite{PFpara,Jermyn,Motruk,Andrei}. The splitting between would-be degenerate energy levels here depends on the system size via a power law. 

\begin{figure}[tbp]
	\begin{center}
	\includegraphics[width=3.3in]{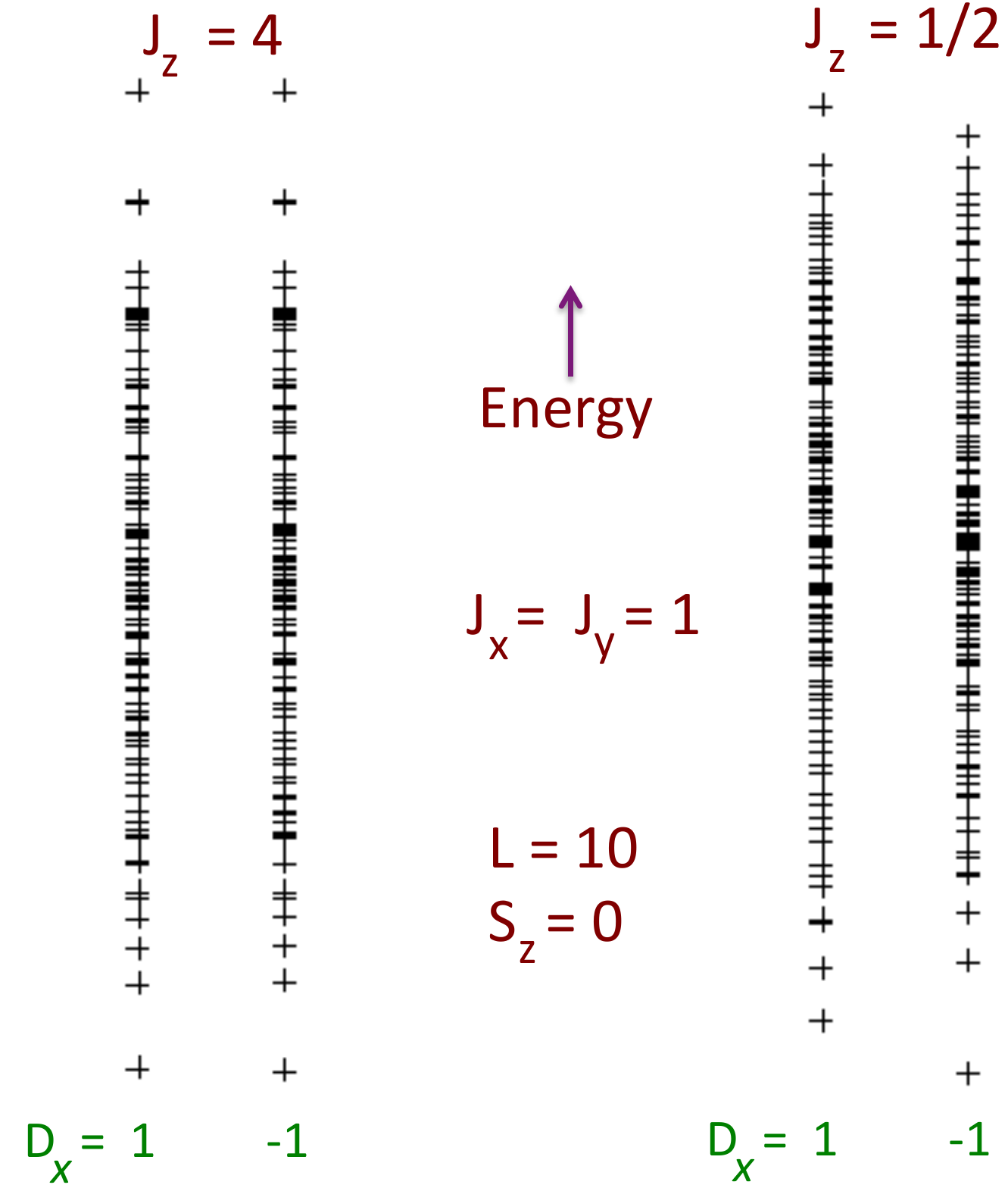}
	\end{center}
	\caption{Spectra of the XXZ spin chain $J_x=J_y=1$ in $D_x=\pm 1$ sectors for 10 sites; on the right the energy is rescaled for easier comparison. In the left plots $J_z=4$ so that the model is ordered, whereas in the right ones $J_z=1/2$ so that it is critical. Only in the ordered case is a pairing between the states apparent. }
	\label{fig:spectrum}
\end{figure}

On the flip side, there is good but not definitive evidence for a strong zero mode once chiral interactions are included in the three-state Potts chain \cite{PFpara,Jermyn}. An operator can be constructed to all orders in perturbation theory that seems to fit the bill \cite{PFpara}, but it has not yet been proven to be normalizable. Moreover, perturbative arguments show that the splitting in pairs of low-lying excited states remains exponentially small in system size \cite{Jermyn}.
In addition, the pairing survives to at least first order in perturbation theory when interactions are included in the Ising/Majorana case  \cite{Kells}. In the case studied in this paper, a glance at the left of figure \ref{fig:spectrum} below provides another compelling suggestion that the pairing survives interactions.

In this paper I promote this suggestion to a proof. By explicit construction I find a strong zero mode in a clean system with strong interactions, the XYZ spin chain with free boundary conditions or equivalently, two Majorana chains coupled by a particular four-fermion interaction. As apparent from  (\ref{Psidef}) below, the operator involves many terms, but the form is quite elegant. I show that is normalizable by using brute force to derive a striking property: its square is the identity operator!  These properties indicate that the strong zero mode is a fundamental property of the system.

\paragraph{The definition of a strong zero mode --}
Given a quantum Hamiltonian $H$ acting on a $L$-site chain with open boundary conditions, 
a strong zero mode $\Psi$ is an operator satisfying \cite{AF}
\begin{itemize}[leftmargin=*]
\item $[H,\Psi]\to 0$ as $L\to\infty$, with finite-size corrections operators whose expectation values vanish as $u^L$ with $|u|<1$.
\item For some $\cal D$ generating a discrete symmetry, $[\Psi, {\cal D}]\ne 0$. 
\item $\Psi^n\propto I$, the identity operator, for some integer $n>1$.
\end{itemize}
Labelling sectors by eigenvalues of ${\cal D}$, the last two conditions mean that acting with $\Psi$ on a state in one sector must give a state in another. Combining this with the first condition means acting with $\Psi$ on an eigenstate of $H$ gives a different state with the same energy, up to corrections going to zero exponentially fast as $L\to\infty$.
The last condition
could be relaxed, but here and elsewhere (e.g.\ for free parafermions \cite{freepara}), it holds. Since this property seems highly non-trivial and fundamental, it seems worthy to emphasize it via this definition. It is worth noting that in general the strong zero mode is not a symmetry -- the spectrum is only asymptotically degenerate unless the finite-size corrections to $[H,\Psi]$ exactly vanish.

\paragraph{The XYZ chain --}

The XYZ chain is an integrable strongly interacting quantum spin chain \cite{Baxbook}.
The Hilbert space is a chain of $L$ two-state systems 
and the Hamiltonian is
\be
H=\sum_{j=1}^{L-1} \left(J_x \sigma^x_j\sigma^x_{j+1} + J_y\sigma^y_j \sigma^y_{j+1} + J_z\sigma^z_j \sigma^z_{j+1}\right)\ ,
\label{Hdef}
\ee
where $\sigma^\alpha_j$ for $\alpha=x,y,z$ is the Pauli matrix $\sigma^\alpha$ acting on the two-state system at the $j$th site. For $J_x=\pm J_y$, this is the XXZ chain, and  $J_x=J_y=J_z$ the antiferromagnetic spin-1/2 Heisenberg chain.
The operators
${D}_\alpha = \prod_{j=1}^L \sigma^\alpha_j $
implement a $\zt\times\zt$ symmetry. When $|J_\alpha|$ is the largest coupling, the model is ordered with the corresponding discrete symmetry spontaneously broken
\cite{Baxbook}. Critical lines separate the different orderings; they occur when $|J_\alpha|=|J_\beta|\ge |J_\gamma|$ for any 
$\alpha\ne \beta\ne \gamma$. 

A Jordan-Wigner transformation maps the Hamiltonian to a fermion one \cite{SML}. Any two of the three terms in (\ref{Hdef}) can be mapped to fermion bilinears, but the third is then a four-fermion term. Thus when one coupling vanishes, the model is free, and turns out to have the same spectrum as two copies of the Ising chain \cite{SML}. The two copies are dual to each other, meaning that when one copy is in the ordered/topological phase, it has a strong edge zero mode \cite{KitMajorana}, while the other is disordered and does not.

As in the Ising/Majorana chain \cite{KitMajorana}, the strong edge zero mode is apparent in a trivially solvable limit. When $J_x=J_y=0$, any basis state $|\zeta\rangle$ in the basis where all $\sigma^z_i$ are diagonal is an eigenstate of the Hamiltonian. The symmetry operator ${D}_x$ then flips all spins, so ${D}_x|\zeta\rangle\ne |\zeta\rangle$. 
Therefore $|\zeta\rangle + D_x |\zeta\rangle$ and $|\zeta\rangle - 
D_x |\zeta\rangle$ are distinct eigenstates of both $D_x$ and $H$ having the same energy. Since the operator $\sigma^z_1$ anticommutes with $D_x$, it must map between the two:  $\sigma_1^z(|\zeta\rangle +D_x |\zeta\rangle) = \pm (|\zeta\rangle -D_x |\zeta\rangle)$. Therefore  in this limit $\Psi^{(0)}=\sigma^z_1$ is a strong zero mode localized at the edge.

Computing the finite size spectra using exact diagonalization strongly suggests that the strong zero mode here survives interactions. The plots for $L=10$ in the XXZ case for two values of $J_z$ are plotted in Fig.\ \ref{fig:spectrum}. The ordered case shows clear evidence for pairing between $D_x=\pm 1$ sectors. Further checks indicate that as $L$ is increased to $14$ (in the sector with vanishing $S_z=\sum_{j=1}^{L} \sigma^z_j$), the splitting between pairs does indeed fall off exponentially. 

\paragraph{The main result --}

Before its proof, I state the main result. 
Letting $X=J_x/J_z,Y=J_y/J_z$ and 
\comment{
With  $J_z$ the largest coupling, the iteration is in powers of $X$ and $Y$, where
\be
X\equiv \frac{J_x}{J_z}\ ,\qquad Y\equiv\frac{J_y}{J_z}\ .
\ee
}
\[
\begin{aligned}
\p{aa'} 
&= Y^{a-a'}(1-X^{-2})\sigma^x_a\sigma^x_{a'} + X^{a-a'}(1-Y^{-2})\sigma^y_a\sigma^y_{a'}\ ,
\end{aligned}
\]
the exact strong edge zero mode for $X^2< 1,Y^2< 1$ is
\be
\Psi = \sum_{S=0}^\infty \sum_{\{a\},b}\,(XY)^{b-1}\sigma^z_{b}\, \prod_{s=1}^S\p{a_{2s-1}a_{2s}}
\label{Psidef}
\ee
where the second sum is over all sets of $2S+1$ positive integers obeying
$
0<a_1<a_2<\dots<a_{2S}< b\ ;
$
if $S=0$ the product is $1$. Terms involving the operator $\sigma^z_{L+1}$ are at least order $L/2$ in $X$ and/or $Y$.

Despite the elegance of this expression, the gargantuan number of terms makes it far from obvious that $\Psi^2$ is proportional to the identity operator. Nevertheless, by brute force I prove in the appendix that for $X^2<1$ and $Y^2<1$,
\be
 \Psi^2 = \frac{1}{(1-X^2)(1-Y^2)}\ .
\label{Psisq}
\ee
The cases $X^2=1$ and/or $Y^2=1$, where the strong zero mode is not normalizable, are critical lines. 

Eigenstate phase transitions occur at the same couplings as the conventional quantum phase transitions.  If a coupling, say $X^2$, is increased past $1$, then $|J_x|$ becomes the largest coupling, and the entire analysis can be repeated with $J_x\leftrightarrow J_z$. A strong zero mode therefore exists everywhere except along the critical lines;
different orderings are associated with different strong zero modes.

\paragraph{Strong zero mode for $J_x=0$ --}

When any of the three couplings vanishes, the XYZ Hamiltonian can be written in terms of fermion bilinears. The strong zero mode is then a linear combination of fermion operators, because commuting a fermion bilinear with any linear combination returns another linear combination. 
Computing the strong zero mode (and all the raising/lowering operators \cite{SML,freepara}) is then rather simple. 

It is not difficult to find this strong zero mode directly, but as a warm-up it is useful to compute it by iteration.
Taking $|J_z|>|J_y|$ and $J_x=0$, the $\zt$ ordering arises from the operator 
$
V\equiv J_z\sum_{j=1}^{L-1} \sigma^z_j\sigma^z_{j+1}
$
dominating $H$. The zeroth order contribution to the strong zero mode, $\Psi^{(0)}=\sigma^z_1$, necessarily commutes with $V$, but not with $H_0=H|_{J_x=0}$:
\[ [H_0^{}\,,\,\sigma^z_1]= J_y [\sigma^y_1\sigma^y_2\,,\,\sigma^z_1] = 2i J_y\sigma^x_1\sigma^y_2\ .\]
The iterative method relies on finding an operator that gives a cancelling contribution when commuted with $V$. Here 
\[ [V,\,\sigma^x_1\sigma^x_2\sigma^z_3]=J_z [\sigma^z_2\sigma^z_3\,,\,\sigma^x_1\sigma^x_2\sigma^z_3]=
2i J_z\sigma^x_1\sigma^y_2\]
yields the first correction to the zero mode, whose commutator with $H_0$ is 
\[ [H_0^{}\,,\,\sigma^z_1-Y\sigma^x_1\sigma^x_2\sigma^z_3]= {-2i}J_yY\sigma^x_1\sigma^x_2\sigma^x_3\sigma^y_4\ .\]
The second correction cancels this term of order $J_y^2$:
\[[V,\, Y^2\sigma^x_1\sigma^x_2\sigma^x_3\sigma^x_4\sigma^z_5] = 2i J_yY\sigma^x_1\sigma^x_2\sigma^x_3\sigma^y_4\ ,
\]
The pattern now is obvious. It is easy to show directly that
\be
\Psi \Big|_{J_x=0} = \sum_{s=0}^\infty\left(-Y\right)^{s}\sigma^z_{2s+1}
\prod_{j=1}^{2s}\sigma^x_j\ 
\label{Psi0}
\ee
obeys
\[\lim_{L\to\infty}[H_0\,,\,\Psi]\Big|_{J_x=0}= 0\ .\]
When $X=0$, the full expression (\ref{Psidef}) reduces to (\ref{Psi0}), which indeed is a sum of Majorana fermion operators.

For the zero mode to be meaningful, it must be normalizable. Because each term in the sum in (\ref{Psi0}) anticommutes with the others, $\Psi^2$ is proportional to the identity:
\[\Psi^2\Big|_{J_x=0} = 1+ Y^2+Y^4 +\dots =(1-Y^2)^{-1}\ ,\]
where the series converges and can be resummed when $Y^2<1$, i.e.\ $|J_y|<|J_z|$. This zero mode therefore exists throughout the $z$-ordered region; only at the critical point $|J_y|=|J_z|$ is it not normalizable. If $|J_z|<|J_y|$ instead, then the entire calculation goes through with $y\leftrightarrow z$. Thus whenever the $J_x=0$ model is ordered, it possesses a strong zero mode.  

\paragraph{Strong zero mode by iteration --}
The iterative approach works for $J_x\ne 0$ as well, although it requires considerably more effort. 
Two useful pieces of notation for this analysis are
\begin{align}
&\q{f(X,Y)}_{aa'}\equiv 2i (f(X,Y)\,\sigma^x_a\sigma^y_{a'} - f(Y,X)\,\sigma^y_a\sigma^x_{a'}),\\
&\s{f(X,Y)}_{aa'} \equiv f(X,Y)\sigma^x_a\sigma^x_{a'}+f(Y,X)\sigma^y_a\sigma^y_{a'}\ ,
\label{double2}
\end{align}
where $f$ is some function of $X$ and $Y$. Any function symmetric in $X$ and $Y$ can be pulled out of or into the double brackets, e.g.\ $\q{XYf(X,Y)}=XY\q{f(X,Y)}$. 

Proceeding as with $J_x=0$, it is easy to verify that
\[
 [H,\,\sigma^z_1]= J_z\q{Y}_{12} = [V,\,\s{Y}_{12}\sigma^z_3]\ .
\]
This yields the first correction to the strong zero mode to be $\Psi^{(1)}= -\s{Y}_{12}\sigma^z_3$. Iterating again yields
\begin{align}
\nonumber
\big[H,\,\sigma^z_1-\s{Y}_{12}\sigma^z_3\big]&=-J_z\s{Y}_{12}\q{Y}_{34} + J_z\q{XY}_{13}\\
&=-\big[V,\widetilde{\Psi}^{(2)}\big]\ ,
\label{v2}
\end{align}
where the putative second-order correction is 
\be
\widetilde{\Psi}^{(2)}=\s{Y}_{12}\s{Y}_{34}\sigma^z_5-\s{XY}_{13}\sigma^z_4\ .
\label{Psi2}
\ee

I say putative because at next order a complication arises. Note that 
\[ \big[H,\, \s{XY}_{13}\sigma^z_4\big] = -J_z\sigma^z_1\q{XY^2}_{23}\sigma^z_4 + \dots \ .
\]
The right-hand-side does appear in a commutator with $V$, but only in combination with another term:
\[
\big[V, \s{f}_{23}(\alpha\sigma^z_1+\beta\sigma^z_4)\big] = J_z(\beta-\alpha)(1-\sigma^z_1\sigma^z_4)\q{f}_{23}\ . 
\]
Thus any commutator of $V$ will yields the product $\q{XY^2}_{23}(1-\sigma^z_1\sigma^z_4)$, which does not arise in $[H,\widetilde{\Psi}^{(2)}]$. The way forward is to note any function of the $\sigma^z_j$ commutes with $V$. Thus at any given order, one can add such a function to the zero mode, and the effects will only be felt at the {\em next} order, after the commutation with the full $H$ is done. Namely, if the second-order correction is modified to be
\[
\Psi^{(2)}=\widetilde{\Psi}^{(2)}+XY\sigma^z_2\ ,
\]
the commutation relation (\ref{v2}) is still satisfied, but $[H,\Psi^{(2)}]$ now gives the desired combination:
\[ \big[H,\,XY\sigma^z_2-\s{XY}_{13}\sigma^z_4\big] = J_z(1-\sigma^z_1\sigma^z_4)\q{XY^2}_{23}+\dots\]
With this modification, a $\Psi^{(3)}$ obeying $[H,\Psi^{(2)}]=-[V,\Psi^{(3)}]$ can indeed be found. 

The iteration can proceed straightforwardly by doing such modifications; it turns out one needs to include a term $(XY)^n\sigma^z_{n+1}$ in $\Psi^{(2n)}$. Intriguingly, another consequence of the modifications is that they make $\Psi^2$ proportional to the identity operator order by order. Although the number of terms grows exponentially, staring at the results for a while leads to (\ref{Psidef}). 


\paragraph{The proof --}

Here I prove that the expression (\ref{Psidef}) satisfies
\be
\lim_{L\to\infty} [H,\Psi] = 0\ .
\label{HPsi}
\ee

For the proof, the XYZ Hamiltonian is split into pieces: 
\[
\begin{aligned}
&H=J_z\sum_{j=1}^{L-1} (T_j+V_j)\\
V_j \equiv \sigma^z_j\sigma^z_{j+1}\ ,&\qquad T_j \equiv X\sigma^x_j\sigma^x_{j+1} + Y\sigma^y_j\sigma^y_{j+1}\ .
\end{aligned}
\]
Each term in the sum (\ref{Psidef}) is labelled as $\psi_{\{a\},b,S}$. 
 It is convenient to separate out of this the pieces $\psi_{\{c\},b,S}^{(q)}$, which has the added restriction that $q<c_1<c_2\dots<c_{2S}$, along with
$\eta_p\equiv\prod_{s=1}^{l} \p{a_{2s-1}\,a_{2s}}$, 
where $a_1<a_2<\dots<a_{2l}<p$.


Finding the appropriate cancellations is best done by ``following'' the $\sigma^z$, i.e.\ keeping track of the location of the the zero or two $\sigma^z_j$ present in each term in $[H,\Psi]$.

First, consider terms in $[H,\Psi]$ with no $\sigma^z_j$, which necessarily involve a non-trivial commutation with the $\sigma^z_b$ in $\Psi$. Three commutators where this happens arise from terms in $\Psi$ 
\[
\begin{aligned} 
{}[T_{b},\eta_b^{}\sigma^z_{b}]&= \eta_b^{}\q{Y}_{b,b+1},\\
XY[T_b, \eta_b^{}\sigma^z_{b+1}]&=-XY\eta_b^{}\q{X}_{b,b+1},\\
(XY)^ 2 [V_{b+1},\eta_b^{}\p{b\,b+1}\sigma^z_{b+2}]&= \eta_b^{}\q{Y(X^2-1)}_{b,b+1}\ .
\end{aligned}
\]
The factors of $XY$ in front are precisely the relative factors appearing in  $\Psi$, so  the corresponding contributions to $[H,\Psi]$  sum to zero for any $\eta_b$. The only other contributions with no $\sigma^z$ arise from the commutators
with $a_{2S}=b-1$:
\begin{eqnarray*}
[T_{b-1}, \eta_p^{}\p{p\,b-1}\sigma^z_b]&=&-\eta_p^{}\q{Y^{p-b+1}(X-X^{-1})}_{pb},\\
XY[V_b, \eta_p^{} \p{pb}\sigma^z_{b+1}]&=& XY\eta_p^{}\q{Y^{p-b}(1-X^{-2})}_{pb}\
\end{eqnarray*}
where $l$ in $\eta_p$ obeys $l=S-1$.
Again, the relative factors of $XY$ are exactly as they occur in the expansion of $\Psi$, and again, the commutators cancel for any $p$ and $\eta_p$. Thus all terms in $[H,\Psi]$ with no explicit $\sigma^z_b$ cancel.

A given term in $[H,\Psi]$ containing $\sigma^z_r\q{pq}$ has an ``inside'' $\sigma^z_r$ when $p<r<q$, and ``outside'' otherwise.  Two canceling pairs with outside $\sigma^z$ are
\[
\begin{aligned}
{}[T_{p},  \p{p\,q}]&=\sigma^z_p\q{X^{p-q+1}(1-Y^{-2})}_{p+1,q}\ ,\\
{} [V_p,  \p{p+1\,q}]&=-\sigma^z_p\q{X^{p-q+1}(1-Y^{-2})}_{p+1,q},\\
{}[T_{q-1},  \p{p\,q}]&=-\sigma^z_{q}\q{Y^{p-q+1}(1-X^{-2})}_{p,q-1}\ ,\\
{} [V_{q-1},  \p{p\,q-1}]&=\sigma^z_{q}\q{Y^{p-q+1}(1-X^{-2})}_{p,q-1}\ ,
\end{aligned}
\]
where $q\ge p+2$.  
Any terms in $[H,\Psi]$ with outside $\sigma^z$  can be obtained by multiplying the above commutators by any $\eta_p\psi^{(q)}$, so all terms with no inside $\sigma^z$ cancel.

The analogous terms with an inside $\sigma^z$ are
\[
\begin{aligned}
{}[T_{p},  \p{p+1\,q}]&=\sigma^z_{p+1}\q{X^{p-q+2}(1-Y^{-2})}_{p,q}\ ,\\
{} [V_p,  \p{p\,q}]&=-\sigma^z_{p+1}\q{X^{p-q}(1-Y^{-2})}_{p,q}\ ,\\
{}[T_{q-1},  \p{p\,q-1}]&=-\sigma^z_{q-1}\q{Y^{p-q+2}(1-X^{-2})}_{p,q}\ ,\\
{}[V_{q-1},  \p{p\,q}]&=\sigma^z_{q-1}\q{Y^{p-q}(1-X^{-2})}_{p,q}\ .
\end{aligned}
\]
The sum ${\cal S}$ of these four commutators is
zero only if $q=p+2$. 
However, when $q>p+2$, an inside $\sigma^z$ arises from
\[
\begin{aligned}
&[T_r,\p{p\,r}\p{r+1\,q}]=(1-X^{-2})(1-Y^{-2})\\
&\ \times \left(\sigma^z_{r+1}\q{Y^{p-r}X^{r-q+2}}_{p,q}
-\sigma^z_r\q{Y^{p-r+1}X^{r-q+1}}_{p,q}\right)\ 
\end{aligned}
\]
for any $p<r<q-1$. Summing these up gives 
$$\sum_{r=p+1}^{q-2} [T_r,\p{p\,r}\p{r+1\,q}] = - {\cal S}\ ,$$
cancelling the original four.
Multiplying these by any $\eta_p\psi^{(q)}$ means that all the commutators with an inside $\sigma_z$ cancel; note that $[V_r,\p{p\,r}\p{r+1\,q}]=0$.

This analysis accounts for all non-vanishing terms in $[H,\Psi]$, and shows all cancel. Thus (\ref{HPsi}) indeed holds. 

\paragraph{$\Psi$ as a symmetry operator}
The boundary conditions used in (\ref{Hdef}) are free at each end. An interesting and useful fact is that if the boundary conditions at the far end are instead fixed, $\Psi$ commutes with $H$ without taking $L\to\infty$. Precisely, the Hamiltonian is modified to $\widetilde{H}=H+J_z V_L$, so that the spin on the site $L+1$ is fixed: no term in $\widetilde{H}$ can flip it, but it still interacts with that at site $L$. Letting $\Psi_L$ be $\Psi$ with the additional requirement that that $b\le L+1$, the preceding proof then shows that $[\widetilde{H},\Psi_L]=0$. 

Thus, with free boundary conditions at one end and fixed at the other, $\Psi_L$ indeed generates a symmetry. The catch is that flipping the spins changes the fixed boundary condition, so $D_x$ no longer generates a symmetry. The pairing becomes trivial: the energies with boundary condition $\sigma^z_{L+1}=1$ are identical to those with $\sigma^z_{L+1}=-1$.
Thus acting with $\Psi_L$ no longer results in degeneracies. 

\paragraph{Finite-Size Effects }

This observation allows the finite-size correction to $[H,\Psi]\to 0$ to be determined. Since $[\widetilde{H},\Psi_L]=0$, it follows that $[H,\Psi_L]=[\widetilde{H}-J_zV_L,\Psi_L]=-J_z[\sigma^z_L\sigma^z_{L+1},\Psi_L]$.  Non-vanishing contributions to this commutator must have $a_{2S}=L$ and $b=L+1$. These are necessarily at least order $L/2$ in $X$ and $Y$, with the leading contributions those from  $X=0$ or $Y=0$, as in (\ref{Psi0}). 
This and the normalizability condition (\ref{Psisq}) ensure that the finite-size corrections do indeed go to zero exponentially fast. 

However, the precise bounding seems difficult to determine. One complication is that $(\Psi_L)^2$ is {\em not} proportional to the identity, although perhaps it could be modified to make it so. Since $\Psi_L$ commutes with $\widetilde{H}$, $\Psi_L^2$ must as well. Therefore, $\Psi_L^2$ may involve powers of $\widetilde{H}$, but since the XYZ chain is integrable, it may involve the commuting higher Hamiltonians as well. Understanding this may provide a route to a much nicer proof of (\ref{Psisq}) than than horrible brute force one in the appendix.

\paragraph{Conclusion--}

I have found the strong zero mode in the clean XYZ chain. An important issue left mostly unexplored is the role the integrability plays. Although very probably the integrability is why the explicit expression is so elegant (and why it could be found straightforwardly by iteration), it is not clear whether or not it is necessary for the strong zero mode to exist. The XYZ chain is not integrable when it has disordered couplings. 
Nonetheless,
low-order calculations indicate that the iterative method still works, but that the expressions become much more complicated. This, along with the work on parafermionic models \cite{PFpara,Jermyn} and on the Ising chain plus interactions \cite{Kells,LY,KF}, gives a good sign that integrability is not necessary for a strong zero mode. 

Moreover, at strong disorder, the resulting MBL phase is argued to also exhibit the same pairing in the spectrum \cite{HNOPS,Bauer,Bahri}. Since MBL phases are believed to resemble integrable systems, for example in having local integrals of motion \cite{SPA,HNO,CKVA}, it would be very interesting if there were a precise correspondence between pairing in clean and very dirty systems. If there is no correspondence, then the interesting question is: what happens at weak disorder?

The eigenstate and quantum phase transitions here occur at the same couplings. The parafermion case, however, hints that this correspondence is not true in general.  
A tool that may prove useful in answering this and other questions is the entanglement spectrum \cite{LH}. This is much easier to compute numerically than the energy spectrum, and it would be interesting to understand if the pairing and the strong zero mode occur in this context as well.

\paragraph{Acknowledgments--} This work was initiated in Santa Barbara whilst I was visiting the Kavli Institute for Theoretical Physics and Microsoft Station Q. I am very grateful to both places for their hospitality. I am very grateful to Fabian Essler for suggesting a look at the XYZ chain, to Erez Berg, Andrei Bernevig, Anushya Chandran, Matthew Fisher, Jack Kemp, Chris Laumann, Chetan Nayak, and Norman Yao for illuminating conversations, and to Jason Alicea, Adam Jermyn and Roger Mong for a stimulating collaboration on \cite{Jermyn}.

\bibliography{XYZzero}

\comment{An indication that the strong zero mode survives interactions may is that the strong zero mode seems connected with topological order, and that often (although not always \cite{FidkowskiKitaev1,FidkowskiKitaev2}) survives. A (more compelling) indication that it may not survive is that the interactions strongly couple the boundary degrees of freedom to the rest, making it much less obvious that it can be modified to commute with an interacting Hamiltonian. }

\comment{\paragraph{Corrections for finite $L$}

Note commutes with fixed b.c on right, i.e.\ we take
\be
H_{\rm free, fixed} = H+ \sigma^z_L\sigma_{L+1}^z\ .
\ee
Then $\Psi$ becomes a symmetry
}

\onecolumngrid
\appendix
\section{Appendix: The proof that $\Psi^2=(1-X^2)^{-1}(1-Y^2)^{-1}$}
Here I prove that for $X<1$, for $Y<1$, $\Psi^2$ obeys (\ref{Psisq}), so that the sum in (\ref{Psidef}) gives a normalizable operator.

\paragraph{Terms that do survive}
Terms in $\Psi^2$  proportional to the identity  come exclusively from the ``diagonal'' terms
$(\psi^{}_{\{a\},b,S})^2$ .
Each diagonal term is built from the square
\begin{eqnarray} \p{aa'}^2 = Y^{2a-2a'}(1-X^{-2})^2+X^{2a-2a'}(1-Y^{-2})^2-\,2\, \sigma^z_a\sigma^z_{a'}\, (XY)^{a-a'}(1-X^{-2})(1-Y^{-2})\ .\label{psq}
\end{eqnarray}
The first aim then is to compute
\be
\sum_{S=0}^\infty\sum_{\{a\},b}(\psi^{}_{\{a\},b,S})^2\Big|_I\ 
=\sum_{S=0}^\infty\sum_{\{a\},b}(XY)^{2b-2}\prod_{j=1}^S (Y^{2a_{2j-1}-2a_{2j}}(1-X^{-2})^2+X^{2a_{2j-1}-2a_{2j}}(1-Y^{-2})^2)\ ,
\label{diagpiece}
\ee
where the $I$ subscript on the left means to include only the terms proportional to the identity.

Despite its nasty looks, this sum turns out to be rather easy to do.
The strategy is to fix a value of $S$, and do the sum over all $a_{1}<a_{2}<\dots <a_{2S}<b$ individually by starting with $b$ and working all the way down. The first is
$$\sum_{b=a_{2S}+1}^\infty (XY)^{2b-2}=\frac{(XY)^{2a_{2S}}}{1-(XY)^2}\ .$$
The factor in the numerator nicely cancels the $X^{-2a_{2S}}$ or $Y^{-2a_{2S}}$ coming from (\ref{diagpiece}). This means that the needed identity for the rest is for a sum over all integers $a_{2j-1}$ and $a_{2j}$ obeying $a_{2j-2}<a_{2j-1}<a_{2j}$ for a fixed $a_{2j-2}$:
\begin{eqnarray*}
\sum_{a_{2j-1}=a_{2j-2}+1}^\infty 
\sum_{a_{2j}=a_{2j-1}+1}^\infty X^{2a_{2j-1}}Y^{2a_{2j}}  = \frac{Y^2}{1-Y^2}
\sum_{a_{2j-1}=a_{2j-2}+1}^\infty (XY)^{2a_{2j-1}} = \frac{(XY)^{2a_{2j-2}+2}}{(Y^{-2}-1)(1-(XY)^2)}\
\end{eqnarray*}
and likewise with $X\leftrightarrow Y$.
Each sum over a pair $a_{2j-1},\,a_{2j}$ in (\ref{diagpiece}) therefore results in a factor $-(XY)^{2a_{2j-2}+2}(X^{-2}+Y^{-2}-2)$, yielding
\begin{eqnarray}
\sum_{\{a\},b,S}(\psi^{}_{\{a\},b,S})^2\Big|_I
=\sum_{S=0}^\infty \frac{(X^2+Y^2-2X^2Y^2)^S}{(1-(XY)^2)^{S+1}}
=\frac{1}{(1-X^2)(1-Y^2)}
\label{diagsum}
\end{eqnarray}
as advertised.
\comment{
\begin{eqnarray*}
\sum_{A<a_{2j-1}<a_{2j}} X^{2a_{2j-1}}Y^{2a_{2j}}  = \sum_{a,a'} X^{2a-2a'} \frac{(XY)^{2a'}}{1-(XY)^2}
=\frac{1}{1-(XY)^2}\sum_{a}X^{2a}\frac{Y^{2a+2}}{1-Y^2}
=\frac{(XY)^{2A+2}}{(1-Y^{-2})(1-(XY)^2)^2}
\end{eqnarray*}}
 
\paragraph{The strategy} 
The much tricker task now is to show that all other terms in $\Psi^2$ cancel. As with the proof that $[H,\Psi]=0$, the key is to organize the terms properly, and to ``follow the $\sigma^z$''. 
$\Psi$ can be written as a sum over monomials in the $\sigma^x_i$ and $\sigma^y_i$ times a single $\sigma^z_b$ at ``the end''. Then, because different Pauli matrices anticommute and same ones square to 1,  the product of any term $\eta\sigma^z_{b}$ with any other one $\eta'\sigma^z_{b'}$ that gives an odd number of $\sigma^z$ must obey 
$$\{\eta\sigma^z_b\,,\,\eta'\sigma^z_{b'}\}\Big|_{{\rm odd}\ \# \sigma^z}=0\ .$$ 
Since all non-diagonal terms in $\Psi^2$ occur as anticommutators of different terms in $\Psi$, 
this vanishing means that all terms in $\Psi^2$ with an odd number of $\sigma^z$ cancel immediately.
Thus the ``follow the $\sigma^z$'' strategy will lead to focussing on the ``last'' two $\sigma^z_b\sigma^z_{b'}$, i.e.\ those where $b'>b>j$ for any other $\sigma^z_j$ present in a given term in $\Psi^2$.
More specifically, I take in turn the cases: 1) terms with only $\sigma^z$ present, 2) terms with no $\sigma^z$ at the end, or two $\sigma^z$ at the end, 3) terms with one or two $\sigma^x$ or $\sigma^y$ in between the last two $\sigma^z$, and 4) terms with any number in between.

\paragraph{1) Terms with only $\sigma^z$ present}
It is simplest to start with the sum over all terms without any $\sigma^x$ or $\sigma^y$ present, namely
\be 
\Psi^2\Big|_{{\rm no}\ \sigma^x,\sigma^y}=\sum_{S=0}^\infty\sum_{\{a\},b,b'} \psi^{}_{\{a\},b,S}\psi^{}_{\{a\},b',S}\ .
\label{noxy}
\ee
The appropriate cancellation can be found simply by examining the last two $\sigma^z$. There are two ways in which these arise here. One is from the explicit $\sigma^z$ in $\Psi$, and come from (\ref{noxy}) with $b\ne b'$. These result in terms in $\Psi^2$ of the form $(XY)^{b+b'-2}\eta^2\sigma^z_b\sigma^z_{b'}.$ The other way is from the product (\ref{psq}) in the diagonal terms in (\ref{noxy}), giving many terms of the form
\[-2(XY)^{b-b'}(1-X^{-2})(1-Y^{-2})\sigma^z_{b}\sigma^z_{b'}\,\eta^2\,(\psi^{(b')}_{\{c\},\tilde{b},S})^2\Big|_{I} ,\]
using the definition after (\ref{HPsi}).  Summing over all $\psi^{(b')}$ here is almost the same calculation as that of (\ref{diagsum}) above; it yields
\[ \sum_{\{c\},\tilde{b},S} (\psi^{(b')}_{\{c\},\tilde{b},S})^2\Big|_{I} = \frac{(XY)^{2b'}}{(1-X^2)(1-Y^2)}\ .\]
Putting these all together that for any fixed $b\ne b'$, the contributions to (\ref{noxy}) cancel:
$$(XY)^{b+b'-2}\eta^2\sigma^z_{b}\sigma^z_{b'}-
(XY)^{b-b'}(XY)^{2b'}(XY)^{-2}\eta^2\sigma^z_{b}\sigma^z_{b'}=0\ .$$
Thus the only terms remaining in (\ref{noxy}) are those diagonal ones proportional to the identity coming from (\ref{diagsum}):
\be
\Psi^2\Big|_{{\rm no}\ \sigma^x,\sigma^y} = \frac{1}{(1-X^2)(1-Y^2)}, 
\label{sumdiag}
\ee
An almost-identical calculation useful below yields
\be
\sum_{\{c\},b,b',S} \psi^{(q)}_{\{c\},b,S} \psi^{(q)}_{\{c\},b',S} = \frac{(XY)^{2q}}{(1-X^2)(1-Y^2)}\ .
\label{sumdiagA}
\ee

\paragraph{2) Terms with no $\sigma^z$ at the end, or two of them}
This calculation is simple to generalize to contributions to $\Psi^2$ with either no $\sigma^z$ at the end, or two of them. These cancel in a very similar way to those discussed in the previous paragraph.
Similarly to $\eta_p$ above, define
$\eta_p'=\prod_{j'=1}^{l'}\p{a'_{2j'-1}a'_{2j'}}$ with $a_{2l'}<p$.
Let $q>a_{2l}$ and consider two terms in $\Psi$:
\be \chi\equiv \eta_q^{}\psi^{(q)}_{\{c\},b,S}\quad{\rm and}\quad \chi'=\eta_p'\p{pq}\psi^{(q)}_{\{c\},b',S}\ ;
\label{etadef}
\ee
the reason for not including $\p{pq}$ inside $\eta'$ will soon become apparent. 
The appearance of $q$ in the definition of $\chi$ is not a typo; it is to ensure that 
there are zero or two $\sigma^z$ at the end in $\chi\chi'$.
Using the identity (\ref{sumdiagA}) to sum over all $\chi\chi'$  gives
\be
\sum_{\{c\},b,b',S}\chi\chi'=
\frac{(XY)^{2q}}{(1-X^2)(1-Y^2)}\eta_q^{}\eta_p'\p{pq}\ .
\label{sumTT}
\ee
and similarly for $\chi'\chi$. 
All other terms in $\Psi^2$ that end in zero or two $\sigma^z$ come from multiplying
$$\xi_t\equiv \eta_q^{}\p{qt}\psi^{(t)}_{\{c\},b,S}\quad{\rm by}\quad \xi'_t=\eta_p'\p{pt}\psi^{(t)}_{\{c\},b',S}\ ,$$
where $t>q>p$. Fixing $t$ momentarily, and using again the sum in (\ref{sumdiagA}) gives
\be
\sum_{\{c\},b,b',S} \xi_t\xi_t'=
\frac{(XY)^{2t}}{(1-X^2)(1-Y^2)}\eta_q^{}\eta_p'\p{qt}\p{pt}\ .
\label{UU}
\ee
There are four terms in the product $\p{qt}\p{pt}$; two with $\sigma^z_t$ and two without, the latter being 
\[
\p{qt}\p{pt}\Big|_{{\rm no}\ \sigma^z_t} = \s{Y^{p+q-2t}(1-X^{-2})^2}_{pq}\ ,
\]
using the notation introduced in (\ref{double2}).
Inserting the no-$\sigma^z$ terms inside (\ref{UU}) gives an operator independent of $r$ up to a constant. 
Doing the sum over $t$ for these terms gives
\[\sum_{t=q+1}^\infty \sum_{\{c\},b,b',S} \xi_t\xi_t'\Big|_{{\rm no}\ \sigma^z_t}=
\frac{1}{(1-X^2)(1-Y^2)}\eta_q^{}\eta_p'\sum_{t=q+1}^\infty\s{Y^{p+q}X^{2t}(1-X^{-2})^2}=
-\frac{(XY)^{2q}}{(1-X^2)(1-Y^2)}\eta_q^{}\eta_p'\p{pq}\ .
\]
Comparing with (\ref{sumTT}) gives
\be
\Psi^2\Big|_{{\rm with\ }\sigma^{x,y};\ {\rm zero\ or\ two\ } \sigma^z\ {\rm at\ end}}=\{\chi,\chi'\} + \sum_t \{\xi_t,\xi'_t\}\Big|_{{\rm no}\ \sigma^z_t}=0.
\ee

\paragraph{3) Terms with $\sigma^z_b\s{F}_{pq}\sigma^z_{b'}$ at the end}
Still unaccounted for are the terms with a single $\sigma^z_{b'}$ at the end, separated from another $\sigma^z_{b}$ by some $\sigma^x$ or $\sigma^y$.  Such terms can come from products like $\chi\chi'$ but where the $q$ superscript in the definition (\ref{etadef}) of $\chi$  is replaced with some value $a_{2l}$ less than $q$. They also can arise allowing the $
\sigma^z_r$ pieces into (\ref{UU}). Two examples of the former type  are
\begin{eqnarray*}
(XY)^3\{\,\sigma^z_1\,,\, \p{23}\sigma^z_4\,\}&=&2(XY)^3\sigma^z_1\sigma_4^z\s{Y(1-X^{-2})}_{23}\ ,\\
(XY)^5\,\{\,\p{12}\sigma^z_3\,,\,\p{13}\sigma^z_4\,\}&=&2(XY)^3\sigma^z_1\sigma^z_4\s{Y(1-X^{-2})(1-Y^{-2})}_{23}\ .
\end{eqnarray*}
An example of the latter type is
\[
\sum_{\{c\},\tilde{b}}\{\,\p{12}\p{34}\psi^{(4)}_{\{c\},\tilde{b}}\,,\,\p{14}\psi^{(4)}_{\{c\},\tilde{b}}\,\}_z=
\frac{(XY)^8}{(1-X^2)(1-Y^2)}\{\p{14}\,,\,\p{12}\p{34}\}_z=-2(XY)^3\sigma^z_1\sigma_4^z\s{Y(1-X^{-2})}_{23}\ ,
\]
where the sum is as in (\ref{sumdiagA}), and the $z$ subscript on the anticommutator means to keep only the terms with $\sigma^z_1\sigma^z_4$. The sum of these three is zero. 

Unfortunately, cancellations in general are not so simple, but the above points the way. Now consider a more general situation, with two $\sigma^x$ or $\sigma^y$ located at $p$ and $q$ in between the last two
$\sigma^z$, located at $b$ and $b'$, with the locations obeying $b<p<q<b'$. The generalizations of the three anticommutators in the previous paragraph are
\begin{align}
\label{three1}
(XY)^{b+b'-2}\{\eta_b^{}\sigma^z_{b},\eta_b'\p{pq}\sigma^z_{b'}\}&={\cal Z}\s{Y^{p-q}(1-X^{-2})}_{pq}\ ,\\
(XY)^{b'+q-2}\{\eta_b^{}\p{bp}\sigma^z_{q},\eta_b'\p{bq}\sigma^z_{b'}\}_z &= {\cal Z}\s{Y^{q-p}(1-X^{-2})(1-Y^{-2})}_{pq}\ ,
\label{three2}\\
\sum\{\eta_b^{}\p{bp}\p{qb'}\psi^{(b')}\,,\,\eta_b'\p{bb'}\psi^{(b')}\}_z
&=-{\cal Z}\s{Y^{q-p}(1-X^{-2})}_{pq}\ ,
\label{three3}
\end{align}
where ${\cal Z}=\{\eta_b,\eta_b'\}(XY)^{b+b'-2}\sigma^z_{b}\sigma^z_{b'}$. 
These three contributions to $\Psi^2$ all end up in the form ${\cal Z}\s{F}_{pq}$, but sum to zero only when $q=p+1$. 

When $q-p>1$, there are other anticommutators that cancel with the preceding three. For example, a contribution similar to (\ref{three2}) is
\[(XY)^{b'+q-2}\{\p{br}\sigma^z_q,\p{bp}\p{rq}\sigma^z_{b'}\}_{{\rm no}\ \sigma^z_r}=-{\cal Z}\s{Y^{p-q}(1-X^{-2})(1-Y^{-2})^2}_{pq}\ .\]
Only terms with no $\sigma^z_{r}$ are included because by assumption the two last $\sigma^z$ are at $b$, $b'$.  It is necessary that $b<p<r<q<b'$; if one instead tried to take $r<p$,  the result would not satisfy the assumption (both parts of the anticommutator would begin with $\p{br}$, and (\ref{psq}) shows that $\p{br}^2$ includes either no $\sigma^z_b$ or the product $\sigma^z_b\sigma^z_{r}$). More generally, there are anticommutators similar to (\ref{three2}) involving multiple $r_n$ obeying $b<p<r_1<r_2<\dots <r_N^{}<q<b'$. For $N$ even and odd these respectively are
\[
\begin{aligned}
(XY)^{b'+q-2}\{\eta_b^{}\p{bp}\p{r_1r_2}\dots\p{r_{N-1}^{}r_N^{}}\sigma^z_q,\eta_b'
\p{br_1}\p{r_2r_3}\dots\p{r_N^{}q}\sigma^z_{b'}\,\}_0^{}&={\cal Z}\s{Y^{q-p}(1-X^{-2})(1-Y^{-2})^{N}}_{pq},\\
(XY)^{b'+q-2}\{\eta_b^{}\p{br_1}\p{r_2r_3}\dots\p{r_{N-1}^{} r_N^{}}\sigma^z_{q},\eta_b'\p{bp}\p{r_1r_2}\dots\p{r_N^{}q}\sigma^z_{b'}\}_0^{}&=-{\cal Z}\s{Y^{q-p}(1-X^{-2})(1-Y^{-2})^{N}}_{pq},
\end{aligned}
\]
where the $0$ subscript means only terms with no $\sigma^z_{r_n}$ are kept.
The key thing to note is that the right-hand side is independent of the specific values of $r_n$, their presence being felt only in the power $(-1)^N(1-Y^{-2})^N$. For each site $R$ in between $p$ and $q$, there are the independent possibilities that $R=r_n$ for some $n$ , contributing a factor $Y^{-2}-1$ to the anticommutator, or $R\ne r_n$ for any $n$, contributing a factor $1$. Thus summing over all possibilities for $r_n$ (including the case $N=0$, the anticommutator (\ref{three2})\,) results in
\be
\sum_{N=0}^{q-p-2}\sum_{\{r\}} (Y^{2}-1)^N=\prod_{N=0}^{q-p-2}(1+Y^{-2}-1) = Y^{2p-2q+2}\ .
\label{sumovers}
\ee
Thus after all the corresponding anticommutators are included, (\ref{three2}) is replaced with
\[
\sum_{\{r\},N}{\cal Z}\s{Y^{q-p}(1-X^{-2})(1-Y^{-2})^{N+1}}_{pq}=
{\cal Z}\s{Y^{p-q+2}(1-X^{-2})(1-Y^{-2})}_{pq}\ .\tag{\ref{three2}'}
\]
Including $r_n$ into (\ref{three3}) in a similar fashion and summing over all possibilities gives
\[
-{\cal Z}\sum_{\{r\},N}\s{Y^{q-p}(1-X^{-2})(1-Y^{-2})^N}_{pq}=
-{\cal Z}\s{Y^{p-q+2}(1-X^{-2})}_{pq}\ .\tag{\ref{three3}'}
\]
Since (\ref{three1}) has an explicit $\sigma^z_b$ on the left-hand side, it is not possible to include any $r_n$ here. The anticommutators in (\ref{three1},\,\ref{three2}',\ref{three3}') therefore account for all contributions to $\Psi^2$ that are of the form ${\cal Z}\s{F(X,Y)}_{pq}$. These three indeed sum to zero for any choice of $\eta_b^{}$, $\eta_b'$. 

This calculation is easy to modify to the case of a single $\sigma^x$ or $\sigma^y$ between $b$ and $b'$, i.e.\ $p<b<q<b'$, $\eta_b\to \eta_p$.  The anticommutator (\ref{three1}) is unchanged if $p\leftrightarrow b$, while the $Y^{q-p}$ in (\ref{three2}) and (\ref{three3}) is modified to $Y^{q+p-2b}$. The $r_n$ here must be modified to have $b<r_1<r_2<\dots<r_m<q$, for the same reason that $r_1>p$ before. This modifies the factor from the sum (\ref{sumovers}) over all $r_n$, to $Y^{2b-2q+2}$. Since $Y^{q+p-2b}Y^{2b-2q+2}=Y^{p-q+2}$, the modifications cancel , and one recovers the same right-hand sides (\ref{three2}',\ref{three3}') even with $p<b$. Thus the cancellation still happens in this case.

\paragraph{4) The remaining terms}

So far I have shown that all terms with $0$, $1$ or $2$ $\sigma^x$ or $\sigma^y$ in between the last two $\sigma^z$ cancel. The remaining terms still comprise the vast majority of those appearing in $\Psi^2$. Luckily, most of these cancel pairwise, and those that do not cancel in a fashion similar to those above.

As above, I start by taking an even number of $\sigma^x$ and $\sigma^y$ to be between the last two $\sigma^z$, at locations $b<p_1<q_1<p_2<q_2<\dots<p_K^{}<q_K^{}<b'$, with $b'$ at the end.
The corresponding terms in $\Psi^2$ involve the operators
$$\sigma^z_{b} \sigma_{p_1}\sigma_{q_1}\sigma_{p_2}\sigma_{q_2}\dots \sigma_{p_K^{}}\sigma_{q_K^{}}\sigma^z_{b'}$$
where a missing superscript in $\sigma$ means it could be either $x$ or $y$. An anticommutator resulting in such terms is the obvious generalization of (\ref{three1}):
\be
{\cal A}=(XY)^{b+b'-2}\{\eta_b^{}\sigma^z_{b},\eta_b'\p{p_1q_1}\dots\p{p_K^{}q_K^{}}\sigma^z_{b'}\}={\cal Z}\prod_{k=1}^K\s{Y^{p_k-q_k}(1-X^{-2})}_{p_kq_k}\ ,
\label{Adef}
\ee
where $b<p_1$.

Many of the others can be seen immediately to cancel. For example, consider one generalization of (\ref{three2}):
\be
\{\eta_b^{}\p{bp_1}\sigma_{\tilde b}\,,\,\eta_b'\p{bq_1}\p{p_2q_2}\dots\p{p_K^{}q_K^{}}\sigma^z
_{b'}\}_z^{}\ ,
\label{gen1}
\ee
where as before the $z$ subscript means to keep only terms containing $\sigma^z_b$. If $\tilde{b}\ne p_k$ or $q_k$ for some $k$, then (\ref{gen1}) violates the stipulation that the last two $\sigma^z$ are at $b$ and $b'$. However, if ${\tilde b}=p_k$ or $\tilde{b}=q_k$ for any $k>1$, then another anticommutator can be built from the same components as (\ref{gen1}) by switching some of them to the other side: 
\be
\{\eta_b^{}\p{bp_1}\p{p_kq_k}\dots\p{p_K^{}q_K^{}}\sigma^z_{b'}\,,\,\eta_b'\p{bq_1}\p{p_2q_2}\dots\p{p_{k-1}q_{k-1}}\sigma^z_{\tilde b}\}_z^{}\ .
\label{gen2}
\ee
This anticommutator (\ref{gen2}) is the opposite of (\ref{gen1}). Thus 
the only possible non-cancelling contributions to $\Psi^2$ of this type occur when $\tilde{b}=q_1$ (by necessity $\tilde{b}>p_1$), thus providing only a very obvious generalization of (\ref{three2}). 

The only contributions that do not cancel in this fashion are necessarily of ``completely overlapping'' type, where it is not possible to switch some components to the other side. For example, (\ref{gen1}) is overlapping when $\tilde{b}=q_1$ because this prohibits $\p{bp_1}$ from being swapped with $\p{bq_1}$. The $r_n$-dependent anticommutators leading up to (\ref{sumovers}) are all overlapping as well. To give more general overlapping contributions, define $M$ distinct integers obeying $1\le t_1<t_2<\dots t_M \le K$. For $M$ odd, let
\begin{align}
{\cal B}_{\{t\}}&=
\p{bp_1} P_{1,t_1}\p{q_{t_1}p_{t_2+1}}P_{t_2+1,t_3}\dots
\p{q_{t_{M-2}}p_{t_{M-1}+1}}P_{t_{M-1}+1,t_{M}}
\sigma^z_{q_{t_M}}\ ,
\nonumber
\\
{\cal C}_{\{t\}}&=
\,\p{bp_{t_1+1}} P_{t_1+1,t_2} \p{q_{t_2}p_{t_3+1}}P_{t_3+1,t_4}\dots 
\p{q_{t_{M-3}}p_{t_{M-2}+1}}P_{t_{M-2}+1,t_{M-1}}
\p{q_{t_{M-1}}q_{t_M}}
\left(\prod_{k=t_M+1}^K\p{p_{k}q_{k}}\right)\sigma^z_{b'} \ .
\nonumber
\end{align}
where $P_{t,t'}=\prod_{k=t}^{t'-1} \p{q_kp_{k+1}}$ with $P_{tt}=1$. For $M$ even, let
\begin{align}
{\cal B}_{\{t\}}&=
\p{bp_1} P_{1,t_1}\p{q_{t_1}p_{t_2+1}}P_{t_2+1,t_3}\dots
\p{q_{t_{M-3}}p_{t_{M-2}+1}}P_{t_{M-2}+1,t_{M-1}}
\p{q_{t_{M-1}}q_{t_M}}
\left(\prod_{k=t_M+1}^K\p{p_{k}q_{k}}\right)\sigma^z_{b'}
\nonumber
\\
{\cal C}_{\{t\}}&=
\,\p{bp_{t_1+1}} P_{t_1+1,t_2} \p{q_{t_2}p_{t_3+1}}P_{t_3+1,t_4}\dots 
\p{q_{t_{M-2}}p_{t_{M-1}+1}}P_{t_{M-1}+1,t_M}
\sigma^z_{q_{t_M}} 
\nonumber.
\end{align}
The complete overlap between ${\cal B}_{\{t\}}$ and ${\cal C}_{\{t\}}$ is because one of each pair $\p{q_{t_{m-1}}p_{t_{m}+1}},\p{q_{t_{m}}p_{t_{m+1}+1}}$ is in ${\cal B}_{\{t\}}$, and the other in ${\cal C}_{\{t\}}$. Because $q_{t_{m-1}}<q_{t_m}<p_{t_{m}+1}<p_{t_{m+1}+1}$, the swap that led to (\ref{gen2}) cancelling with (\ref{gen1}) is not possible here.
The product of the two for any $M$ therefore is given by
\be{\cal B}_{\{t\}}{\cal C}_{\{t\}}=(-1)^M\p{bp_1}\p{q_{t_{M-1}}q_{t_M}}
\left(\prod_{m=0}^{M-2}\p{q_{t_{m}}p_{t_{m+1}+1}}\right)\left(\prod_{m=0}^{M-1}P_{t_m+1,t_{m+1}}\right)\left(\prod_{k=t_M+1}^K\p{p_{k}q_{k}}\right)\sigma^z_{q_{t_M}}\sigma^z_{b'}
\label{BC}
\ee
where $t_0\equiv 0$ and $q_0\equiv b$; for $M=1$ the product ending at $M-2$ is omitted. 
The sign in front arises because for odd $M$ the end $\sigma^z_{q_{t_M}}$ appears in ${\cal B}_{\{t\}}$ while  $\p{q_{t_{M-1}}q_{t_M}}$ appears in ${\cal C}_{\{t\}}$, and vice versa for even $M$. The two operators anticommute, so writing  ${\cal B}_{\{t\}}{\cal C}_{\{t\}}$ in the form (\ref{BC}) results in the $(-1)^M$. 

In the special case $t_M=q_K$, there is another contribution to $\Psi^2$ generalizing (\ref{three3}), i.e.\ where the $\sigma^z_{b'}$ is obtained by a $\sigma^x_{b'}$ in $\tilde{\cal B}$ and a $\sigma^y_{b'}$ in $\tilde{\cal C}$, or vice versa.  A slight extension of the above calculation gives one additional contribution for each $\{\tilde{t}\}=\{t_1,t_2,\dots t_{M-1}^{},q_K\}$, which for even $M$ is
\be
\tilde{\cal B}_{\{\tilde{t}\}}\tilde{\cal C}_{\{\tilde{t}\}}
=\p{bp_1}\p{q_{t_{M-1}}b'}\p{q_Kb'}
\left(\prod_{m=0}^{M-2}\p{q_{t_{m}}p_{t_{m+1}+1}}\right)\left(\prod_{m=0}^{M-1}P_{t_m+1,t_{m+1}}\right)\psi^{(b')}\psi^{(b')}
\label{BCtilde}
\ee
while for odd $M$ the $\p{q_{t_{M-1}}b'}$ and $\p{q_Kb'}$ are reversed.

Each $p_k$ and $q_k$ appears precisely once in ${\cal B}_{\{t\}}{\cal C}_{\{t\}}$, as apparent from the explicit expression (\ref{BC}). This is true for any sequence $\{t\}$; its role is solely to label how the overlaps happen.  Therefore terms with different $\{t\}$ can cancel with other as well as with ${\cal A}$, but this cancellation is subtler than those discussed above: one needs to ``split'' apart the overlapping bits. To this end, let $\p{aa'}=\p{aa'}_x+\p{aa'}_y$, where
$$\p{aa'}_x=Y^{a-a'}(1-X^{-2})\sigma^x_a\sigma^x_{a'}\quad,\quad
\p{aa'}_y=X^{a-a'}(1-Y^{-2})\sigma^x_a\sigma^x_{a'}\ .
$$ %
The key observation is then that 
\be
\begin{aligned}
\p{q_{t_{m-1}}p_{t_{m}+1}}_x^{}\p{q_{t_{m}}p_{t_{m+1}+1}}_x^{}&=
\p{q_{t_{m-1}}p_{t_{m+1}+1}}_x^{}\p{q_{t_{m}}p_{t_m+1}}_x^{}\ ,\\
\p{q_{t_{m-1}}p_{t_{m}+1}}_y^{}\p{q_{t_{m}}p_{t_{m+1}+1}}_y^{}&=
\p{q_{t_{m-1}}p_{t_{m+1}+1}}_y^{}\p{q_{t_{m}}p_{t_m+1}}_y^{}\ .
\end{aligned}
\label{id1}
\ee
Consider then a sequence $\{\hat t\}$, which is $\{t\}$ with $t_m$ removed, so it has only $M-1$ entries. Then for $1\le m<M-1$
\be
\frac{{\cal B}_{\{t\}}{\cal C}_{\{t\}}}{{\cal B}_{\{\hat t\}}{\cal C}_{\{\hat t\}}}
=\frac{(-1)^M\p{q_{t_{m-1}}p_{t_{m}+1}}P_{t_{m-1}+1,t_{m}}\p{q_{t_{m}}p_{t_{m+1}+1}}P_{t_m+1,t_{m+1}} }
{(-1)^{M-1}\p{q_{t_{m-1}}p_{t_{m+1}+1}}P_{t_{m-1}+1,t_{m+1}}}
= -\frac{\p{q_{t_{m-1}}p_{t_{m}+1}}\p{q_{t_{m}}p_{t_{m+1}+1}} }{\p{q_{t_{m-1}}p_{t_{m+1}+1}}\p{q_{t_{m}}p_{t_m+1}}}.
\label{id2}
\ee
For $m=M-1$, the relation is modified by sending $p_{t_{M}+1}\to q_{t_M}$.
The ratio in (\ref{id2}) is exactly that for which the identity in (\ref{id1}) holds. Thus {\em any} time $\p{q_{t_{m-1}}p_{t_{m}+1}}_x^{}$ is followed by 
$\p{q_{t_{m}}p_{t_{m+1}+1}}_x^{}$, these terms in ${\cal B}_{\{t\}}{\cal C}_{\{t\}}$ are cancelled by the corresponding terms in ${\cal B}_{\{\hat t\}}{\cal C}_{\{\hat t\}}$. The same holds for $y$ subscripts. Since $\Psi^2$ requires a sum over all sequences $\{t\}$, this cancels almost everything. The only ones which survive are when $\p{q_{t_{m-1}}p_{t_{m}+1}}_x^{}$ is followed by $\p{q_{t_{m}}p_{t_{m+1}+1}}_y^{}$ and likewise with the subscripts $x\leftrightarrow y$. Moreover, within the product $P_{t_m+1,t_{m+1}}$, only the $y$ components survive when $\p{q_{t_{m}}p_{t_{m+1}+1}}_x^{}$ survives, and likewise with the subscripts $x\leftrightarrow y$. Explicitly, what survives for example is proportional to
\[
\dots\sigma^y_{q_{t_m}}\sigma^x_{p_{t_m}+1}(\sigma^x_{q_{t_m+1}}\dots \sigma^x_{p_{t_{m+1}}})\sigma^x_{q_{t_{m+1}}} \sigma^y_{p_{t_{m+1}+1}}(\sigma^y_{q_{t_{m+1}+1}}\dots \sigma^y_{q_{t_{m+2}}})\dots
\]where the parentheses enclose the terms coming from $P_{t_m+1,t_{m+1}}$ restricted to all $\sigma^x$
and $P_{t_{m+1}+1,t_{m+2}}$ restricted to all $\sigma^y$, and the others come from $\p{q_{t_{m-1}}p_{t_{m}+1}}_x$, $\p{q_{t_m}p_{t_{m+1}+1}}_y$,  and $\p{q_{t_{m+1}}p_{t_{m+2}+1}}_x$. This pattern even persists at $t_M$, because multiplying by the explicit $\sigma^z_{q_{t_M}}$ changes $\sigma^x_{q_{t_M}}\leftrightarrow \sigma^y_{q_{t_M}}$, so that it coincides with the superscript at $p_{t_M}$ coming from $P_{t_{M-1}+1,t_M}$.

The upshot is that of the $2^{t_M}$ terms for a given ${\cal B}_{\{t\}}{\cal C}_{\{t\}}$ only two of them for each $\{t\}$ survive the sum over $\{t\}$. The two that survive are quite simple to characterize: each $t_m$ labels a ``domain wall'' between $q_{t_{m}}$ and $p_{t_m+1}$ separating regions of $\sigma^x$ and $\sigma^y$. For values of $k$ greater than $t_M$, domain walls can occur between any $q_k$ and $p_{k+1}$, as follows from the product over $k$ in (\ref{BC}). 
Note each of the $2^K$ terms coming from splitting apart ${\cal A}$ in (\ref{Adef}) has exactly the same possibilities for domain walls! Domain walls in the surviving terms always come between a $q_k$ and the subsequent $p_{k+1}$, never between $p_k$ and $q_k$. Terms obtained by including the $r_n$ as above also behave in the same way, since by construction including any $r_n$ does not change the resulting operator, only the coefficient.  Moreover, the coefficient of each term can be obtained by multiplying bits coming from each $(p_k,q_k)$. Therefore, all surviving contributions can be written as a sum over terms of a remarkably simple form. Namely,  putting back in the $\eta_b^{}$ and $\eta_b'$ to account for operators before the last two $\sigma^z$ means that
\be
\sum_{\{t\}}\{ \eta_b^{} {\cal B}_{\{t\}}\,,\, \eta_b'{\cal C}_{\{t\}}\}_z^{} = 
{\cal Z} \sum_{\{\alpha_k=x,y\}}
\prod_{k=1}^K \left(F_{k,\alpha_k}^{}\,\sigma^{\alpha_k}_{p_k}\sigma^{\alpha_k}_{q_k}\right).
\label{Fdef}
\ee
The sum is over all all choices of $\alpha_k=x,y$.
By symmetry under $x\leftrightarrow y$, the functions obey $F_{k,x}(X,Y)=F_{k,y}(Y,X)$. 

The remaining exercise is therefore to compute the coefficients $F_{k,\alpha_k}$ for all the remaining overlapping contributions. After all this work, this is relatively simple. First consider ${\cal A}$ coming from (\ref{Adef}), which by definition of $\p{p_kq_k}$ yields
\be
{\cal A}={\cal Z}\sum_{\{\alpha_k=x,y\}}
\prod_{k=1}^K \left(F^{({\cal A})}_{k,\alpha_k}\sigma^{\alpha_k}_{p_k}\sigma^{\alpha_k}_{q_k}\right)\qquad\hbox { with }\qquad F^{({\cal A})}_{k,x}=Y^{p_k-q_k}(1-X^{-2})\ .
\label{FAdef}
\ee
If $k>t_{M}$, then the only appearance of $p_k$ and $q_k$ in (\ref{Fdef}) is via the product at the end of (\ref{BC}), so $F_{k>t_k,x}=Y^{p_k-q_k}(1-X^{-2})$, as in $F^{({\cal A})}_{k,x}$. 

If $k<t_M$, there are two contributions to $F_{k,x}$. One is from the explicit factor $\p{q_{k}p_{k+1}}$ if $k\ne t_{m}$ or  $\p{q_{t_{m-1}}p_{t_m+1}}$ when $k=t_m$. In either case, it contributes a factor $Y^{q_k-p_k}(1-X^{-2})$. The second contribution comes from including and summing over the $r_n$. In the case $M=K=1$ discussed above (\ref{three2}'), the $r_n$ can be included only between $p_1$ and $q_1$. It is easy to see that the analogous result for general $M\le K$ is that the $r_n$ can only be included between $p_k$ and $q_k$ for all $k\le t_M$. Attempting to include an $r_n$ in between $q_k$ and $p_{k+1}$ wrecks the overlapping, and so the contributions cancel as in (\ref{gen1},\ref{gen2}). Thus the sum over $r_n$ between $p_k$ and $q_k$ goes as in (\ref{sumovers}), contributing a multiplicative factor 
$Y^{2+2p_k-2q_k}$ to $F_{k,x}$. Combining the two factors gives therefore
\[F_{k<t_M,x}=Y^{q_k-p_k}(1-X^{-2})Y^{2p_k-2q_k+2} = Y^{p_k-q_k+2}(1-X^{-2})\ ,\]
i.e.\ the result for $k>t_M$ multiplied by $Y^2$.

As is obvious from (\ref{BC}), the case $k=t_M$ looks different. For example, $\p{q_{t_{M-1}q_{t_M}}}_y\sigma^z_{t_{q_M}}$ contributes $iX^{-q_{t_M}}(1-Y^{-2})$ to $F_{t_M,x}$. 
It is also convenient to absorb into this factor the other constants from the terms in (\ref{BC}) not included into the above. Assembling all  them all gives
\[F_{t_M,x}=(-1)^{M-1}(XY)^{q_{t_{M}}}Y^{-p_{t_M}}X^{-q_{t_M}}(1-X^{-2})(1-Y^{-2})Y^{2p_{t_M}-2q_{t_M}+2} = (-1)^{M-1}Y^{p_{t_M}-q_{t_M}}(Y^2-1)
\]
again very similar to the other cases. This indeed agrees with the $M=K=1$ case (\ref{three2}'). Putting these expressions together gives
\be
F_{k,x}= Y^{p_k-q_k}(1-X^{-2})\times 
\begin{cases}
Y^2\quad&\hbox{ for }k<t_M\\
(Y^2-1)&\hbox{ for }k=t_M\\
1\quad&\hbox{ for }k>t_M
\end{cases}
\label{FF}
\ee
The $(-1)^M$ in (\ref{BC}) cancels, because $\alpha_{t_M}=x$ when $M$ is odd and $\alpha_{t_M}=y$ when $M$ is even, and multiplying by $\sigma_{q_{t_M}}$ gives different signs in the two cases.
The computation of the analogous coefficient coming from the last piece (\ref{BCtilde}) is very similar to the preceding, yielding
\be
\tilde{F}_{k,x}= Y^{p_k-q_k}(1-X^{-2})\times 
\begin{cases}
Y^2\quad&\hbox{ for }k<t_M=K\\
-Y^2&\hbox{ for }k=t_M=K
\end{cases}
\label{Ft}
\ee
in agreement with (\ref{three3}') for $M=K=1$. 

The completion of the proof is just a few sums away.  For simplicity, first consider the case of no domain walls, i.e.\ either $\alpha_k=x$ for all $k$ or $\alpha_k=y$ for all $k$. There are still multiple contributions, since $t_1=t_M$ ranges from $1\dots K$, each of which contributes 
\[\prod_{k=1}^K F_{k,x}\Big|_{M=1}=Y^{\sum_k(p_k-q_k)}(1-X^{-2})^K (Y^2-1)Y^{2t_1-2}\ .\]
Summing these along with those from the two other contributions (\ref{FAdef}) and (\ref{Ft}) gives
\[1+ \sum_{t_1=1}^K Y^{2t_1-2}(Y^2-1) -Y^{2K}= 1-(1-Y^{2K})-Y^{2K}=0\ ,\]
where the common factor $Y^{\sum_k(p_k-q_k)}(1-X^{-2})^K$ is omitted.
These contributions indeed cancel!

In general, let $d_l$ be the locations of the domain walls. The first $M-1$ domain walls necessary occur at the $t_m$, so $d_m=t_m$ for $m\le M-1$. There need not be a domain wall at $t_M$,  so for fixed number of domain walls $D$, $M\le D+1$.  Thus to find all the contributions for a given $d_1,d_2,\dots d_D$, one first fixes $M$ and sums over all $t_M$ allowed, and then sums over all $M\le D+1$. The sum over $t_M$ goes from $d_{M-1}+1$ to $d_M$,  (or to $K$, if there are only $M-1$ domain walls). Again ignoring common factors, the sum over all terms for $D$ even is
\be
\begin{aligned}
&1 + (Y^2-1)\sum_{t_1=1}^{d_1}Y^{2t_1-2}\ +\ Y^{2d_1}(X^2-1)\sum_{t_2=d_1+1}^{d_2} X^{2(t_2-d_1-1)}
\ +\ Y^{2d_1}X^{2(d_2-d_1)} (Y^2-1)\sum_{t_3=d_2+1}^{d_3}Y^{2(t_3-d_2-1)}\\
&\qquad\qquad +\ \dots\ +\ Y^{2d_1}X^{2(d_2-d_1)}\cdots
Y^{2(d_{D}-d_{D-1})}\left((X^2-1)\sum_{t_M=d_{D}+1}^K X^{2(t_M-d_{D}-1)} -X^{2(K-d_D)}\right)
\end{aligned}
\label{finally}
\ee
where the contributions are respectively from (\ref{FAdef})  (effectively $M=0$), $M=1$, $M=2$, $M=3$ and, within the last parentheses, from $M=D+1$ and from (\ref{Ft}). Doing the sums over the $t_M$ in (\ref{finally}) gives
\[
\begin{aligned}
&1 + (Y^{2d_1}-1)+Y^{2d_1}(X^{2(d_2-d_1)}-1) + Y^{2d_1}X^{2(d_2-d_1)}(Y^{2(d_3-d_2)}-1)\\ 
&\qquad\qquad \qquad\qquad\qquad\qquad+\ \dots\ +\ Y^{2d_1}X^{2(d_2-d_1)}\cdots  Y^{2(d_{D}-d_{D-1})}\left((X^{2(K-d_{D})} -1)-X^{2(K-d_D)}\right)
\\
&= 0.
\end{aligned}
\]
 For $D$ odd, the sum is virtually identical; just send $X\to Y$ inside the last parentheses, and $Y\to X$ in the last factor before that. It thus is also zero.
 
The last remaining case to check is when there are an odd number of $\sigma^x,\sigma^y$ in between the last two $\sigma^z$ in $[H,\Psi]$. This requires only a slight modification of the preceding few pages, exactly as described at the end of case 3) above.
 
This horrible brute force proof is done! The huge number of cancellations are a pretty strong indication that 
there is a much simpler method, but I haven't found it. A guess for how this might work is presented in the main text.
\end{document}